\documentclass[%
 reprint,
 amsmath,amssymb,
 aps,
 prl,
 longbibliography,
 lengthcheck,%
]{revtex4-1}

\usepackage{graphicx}
\usepackage{dcolumn}
\usepackage{bm}
\usepackage{hyperref}

\begin{document}

\title{Estimating Total Solar Irradiance during the 21st century}
\author{  Victor Manuel Velasco Herrera,$^1$ Blanca Mendoza$^{1}$ and Graciela Velasco Herrera.$^{2}$} 
\email{Electronic address: vmv@geofisica.unam.mx  \\}

\affiliation{$^1$Departamento de Ciencias Espaciales, Instituto de Geof\'isica, Universidad Nacional Aut\' onoma de M\'exico, Ciudad Universitaria, Coyoac\'an, 04510, M\'exico D.F., M\'EXICO}
\affiliation{ $^2$Centro de Ciencias Aplicadas y Desarrollo Tecnol\' ogico, Universidad Nacional Aut\'onoma de M\'exico}

\begin{abstract}
The reconstruction and prediction of solar activity is one of the current problems in dynamo theory and global climate modeling. We estimate the Total Solar Irradiance for the next hundred years based on the Least Square Support Vector Machine. We found that the  next secular solar minimum will occur between the years $2003$ and $2063$ with an average of $1365.4$ $W/m^2$ close to the Dalton or Modern minima. We calculate the radiative forcing between the modern maximum and the  $21^{st}$  century minimum to be $-0.1$$ W/m^2$.
\end{abstract}

\pacs{Valid PACS appear here}
\maketitle


The  modern comprehension of solar variability possibly began when Schwabe published the periodicity of sunspots in $1843$ \cite{ref1}. In $1894$, Maunder published a discovery that has maintained the Solar Physics in an impasse. In his famous work on ``A Prolonged Sunspot Minimum'' Maunder wrote \cite{ref2}: 
\\ \\
``The sequence of maximum and minimum has, in fact, been unfailing during the present century [$ . . $] and yet there [$ . . $], the ordinary solar cycle was once interrupted, and one long period of almost unbroken quiescence prevailed''\\

The Maunder's discovery went unnoticed and forgotten until $1976$ when Eddy brought it to light again \cite{ref3}. The existence of prolonged solar minima has been one of the most controversial questions in Solar Physics. However the possibility of prediction of  new periods of diminished solar activity is even more controversial.

To estimate future  solar activity, several methods have been used, for instance dynamo models, spectral methods, regression methods or neuronal network methods \cite{ref4}. These estimations have in common that they are applied to short-time reconstructed series and that they do not discuss the relative accuracy of the methods.

The fluctuations of the solar time series is a tool that helps to study the solar magnetic field as well as to understand the solar dynamo. These fluctuations can occur for instance in the amplitude, phase, frequency, energy and power of the solar phenomena.

The majority of solar activity analysis focuses  on fluctuations of the amplitude. In this paper, we propose to consider not only the fluctuations  in amplitude but also  in the power of the Total Solar Irradiance (TSI)  as a physical measure of the energy released by the solar dynamo, which contributes  to understand  the nature of `profound solar magnetic field in calm".
Regardless of the mechanism that produces solar activity minima (stochastic, chaotic, intermittent or quasi-periodical processes), the study of these minima is very important for the solar dynamo theory, as well as for its  impact on  solar-terrestrial relationships \cite{ref5,ref6}.

Recent studies suggest that the mid-term ($1- 2$ years) and the secular periodicities are the product of chaotic quasi-periodic processes and not of stochastic processes or intermittent process \cite{ref7,ref8}. Different spectral analysis of solar activity series \cite{ref8, ref9} show several significant long-term periodicities. It is also known that the solar cycle (Schwabe periodicity \cite{ref1}) varies cyclically with a mean period of about 11-years and the magnetic cycle (Hale cycle \cite{ref10}) with a mean period of about of 22-years. This behavior  motivates attempts to predict solar activity, especially now that an unexpectedly low activity solar cycle 23 occurred and could be the sign of the beginning of a new secular solar minimum \cite{ref8}. The behavior of the solar cycle $23$ minimum has shown an activity decline not previously seen in the past cycles for which spatial observations exist \cite{ref11,ref12,ref13,ref14}. 

The descending phase and minimum measurements  of solar cycle 23 show that in particular the TSI has fallen below the  previous two solar minima values: the mean PMOD composite TSI  for September  $2008$ is $1365.26 \pm 0.16$ $Wm^{-2}$ , compared to  $1365.45$  $Wm^{-2}$  in $1996$ or $1365.57$ $Wm^{-2}$  in $1986$ \cite{ref15}. 

Over the $11$-years solar cycle, TSI variations of  $\sim$$0.07 \%$ have been observed between solar minimum and maximum \cite{ref16}. This modulation is mainly due to the interplay between dark sunspots and bright faculae and network elements\cite{ref17}. Studies using cosmogenic isotope data and sunspot data \cite{ref8,ref18} indicate that currently we are within a grand activity maximum which began  after  $\sim$$1930$. 

Studying the solar wind, the interplanetary magnetic field strength and the open solar flux over the past century, Lockwood \cite{ref19} found that all three parameters show a long-term rise peaking around $1955$ and $1986$ and then decline, yielding predictions that the grand maximum will end in the years $2013$, $2014$ or $2027$ depending on the parameter used.   

Other  works  indicate that the current maximum will not last longer than two or three solar cycles more \cite{ref20}. Furthermore, it has been suggested that  a  Dalton-type minimum has already began in solar cycle $23$  reaching \cite{ref21,ref22}  to  solar cycles $24$ and $25$, while the Solar Cycle $24$ Prediction Panel  indicates a lower limit of $90\pm10$ for the maximum sunspot number  \cite{ref23}  of solar cycle $24$.

\begin{figure}[b]
\includegraphics [width=0.47\textwidth]{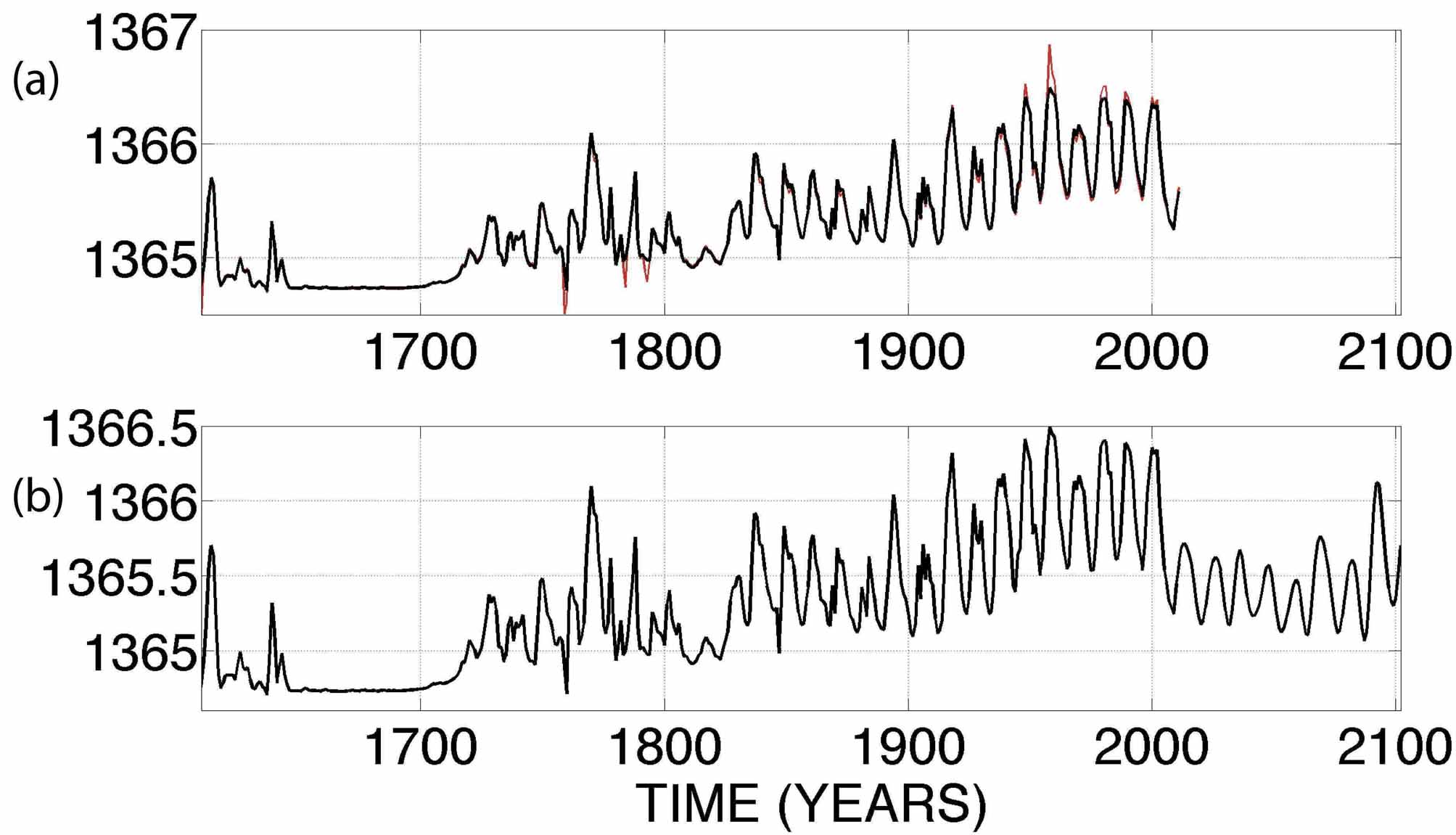}
\caption{\label{fig:epsart} 
Time series. (a) The TSI-KRIVOVA-PMOD  (red line)  from $1610$ to $2010$  superposed on  the times series estimated with the Least Squares Support Vector Machines (LS-SVM) between $1610$ and $2010$ (black line).  (b) The LS-SVM model time series (black line)  between $1610$ and $2100$ }
\end{figure}

To project the TSI for the next hundred years, we use a method based on Least Squares Support Vector Machines (LS-SVM) with Nonlinear Autoregressive Exogenous (NARX) model  and with radial basis function (RBF) kernel that  allows a better  precision in the estimation of the future values  of a time series \cite{ref24,ref25}.

\begin{figure}
\center
 \noindent\includegraphics[width=0.47\textwidth]{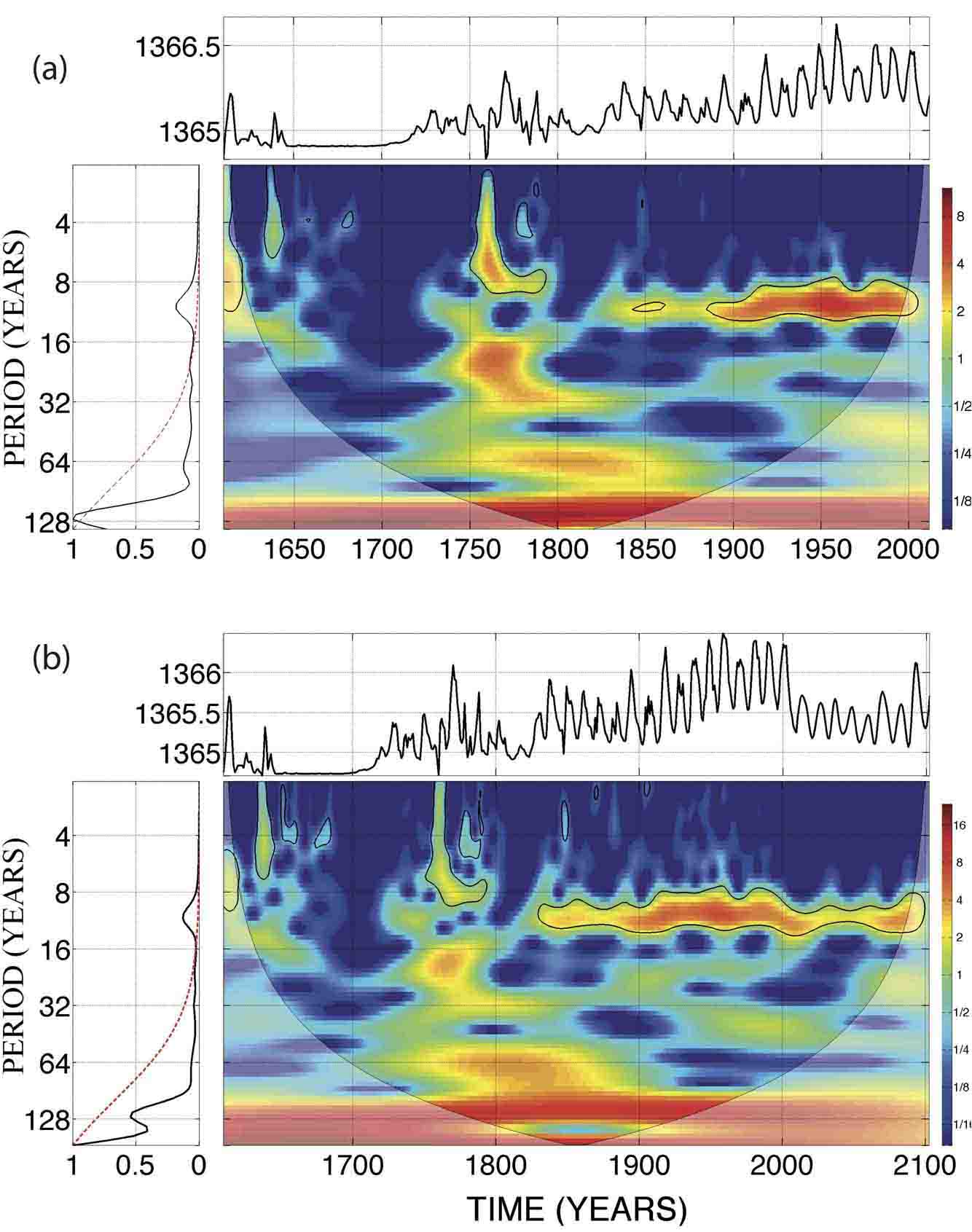}
 \caption{(a) Upper panel, TSI-KRIVOVA-PMOD; right panel spectral wavelet analysis; left panel global wavelet. (b) The same as in (a) but for the  LS-SVM model. }
 \end{figure}

Since the instrumental TSI records began in $1979$, we work with reconstructed  TSI  series. Long-term reconstructions of TSI \cite{ref26,ref27} show epochs of maxima and minima when substantial changes in the TSI occur. These changes can contribute to climate variability \cite{ref28} and  new estimate of the TSI for the 21st century, will have an important impact on climate modeling. 

We tested three TSI recent reconstructions. Two of them use a model based on the evolution of the Sun«s total and open magnetic flux: The Wang reconstruction \cite{ref29} consider differential rotation, supergranular convection and meridional flow; this model is used to derive two TSI reconstructions, one with and one without a secularly varying background; from $1850$ onwards these reconstructions are the recommended solar forcings for the fifth Coupled Model Intercomparison Project $20$th century  simulations \cite{ref30}.  

The Krivova reconstruction \cite{ref31} relies on time constants representing the decay and conversion of the different photospheric magnetic flux components. The  Stenhilber reconstruction\cite{ref26}, uses the observationally derived relationship between TSI and the open solar magnetic field, the latter obtained from the cosmogenic radionuclide $^{10}Be$.  

However, the best reconstruction to apply our method is the Krivova reconstruction \cite{ref31} because it has a better temporal and spectral resolution compared to the Stenhilber reconstruction \cite{ref26}  and the correlation between the reconstructed and the PMOD composite is better than the Wang reconstruction\cite{ref29}. 

From $1610$ to $1975$ we use the Krivova reconstruction \cite{ref31}  and from $1976$ to $2010$ we use the PMOD composite. In this work we use  the TSI-KRIVOVA-PMOD between 1610 and 2010.

To search for the future TSI values, the NARX LS-SVM was trained with  $80\%$ of  random data of the TSI-KRIVOVA-PMOD, obtaining a mean squared error (MSE) of $0.0084$, the testing of the remaining  $20\%$ presents an MSE of $0.0073$. 

We plot the TSI-KRIVOVA-PMOD with red line and the NARX LS-SVM model with black line in Fig. $1a$. It is clear that the  NARX LS-SVM model (black line) reproduces very well the TSI-KRIVOVA-PMOD; the linear correlation coefficient between the TSI-KRIVOVA and the NARX LS-SVM model series for $1610$-$1975$ is $r = 0.9969$, while $r$ value for the PMOD and the NARX LS-SVM model for $1976$-$2010$ interval is $0.9959$.

In  Fig. $1b$ we present the NARX LS-SVM  future TSI estimation (black line) between $2011$ and $2100$ with standard deviation $\sigma= 0.4318$ $Wm^{-2}$. We notice a decreasing trend of the TSI  between $2003$ and  $2063$,  coinciding with other types of prediction adopting different methods\cite{ref8,ref19,ref20,ref22}.

However, it is not enough to conclude that the amplitudes among the reconstructed, the composite and the modeled time series are similar, it is also necessary to compare the spectral characteristics. 

We apply the wavelet analysis using the Morlet function \cite{ref32}, to quantify the TSI time series till 2100 A.D. and to analyse local variations of multiple periodicities. This method   provides a higher resolution of periodicity, allows us to calculate the phase, and  to filter the TSI in  bandwidths \cite{ref33}. 

Also, to calculate the confidence level we used the normalizepdf function, in this way the   TSI will have a gaussian distribution\cite{ref34}. Wavelet meaningful periodicities (confidence level greater than $95\%$) must be inside the cone of influence (COI), which is the region of the wavelet spectrum outside which the edge effects become important \cite{ref35}. We also include the global spectra in the wavelet plots to show the power contribution of each periodicity inside the COI \cite{ref36}. 

We established our significance levels in the global wavelet spectra with a simple red noise model (increasing power with decreasing frequency \cite{ref37}). We only took into account those periodicities above the red noise level.

In Fig. $2$ we show the  wavelet analysis  for  the  TSI-KRIVOVA-PMOD (Fig. $2a$) and the NARX LS-SVM model (Fig. $2b$). In the central panel of Fig. $2a$ the $11$-years cycle appears attenuated during the secular minima and is stronger since $\sim$$1875$, while the $\sim$$120$-years cycle keeps a   more or less uniform power along all the time-span considered, both periodicities appear above the red noise level in the global spectra (right panel). Fig. $2b$ shows roughly the same spectral evolution of Fig. $2a$  but the two peaks above the red noise level are  the $11$-years and the $240$-years. Also Fig. $2b$ indicates that the  $11$-years cycle will be attenuated  during the next $60$ years (to $2063$), which is a characteristic of a secular minima.

In Fig. $3$ we show a further analysis of the amplitudes (black areas) and phases (blue lines), that allow us to quantify the starting and ending of a cycle  of the NARX LS-SVM model and were obtained using the inverse wavelet transform \cite{ref32}. Fig. $3b$ shows the $11$-years cycle, as sunspots were so scarce during the ``Prolonged Sunspot Minimum'', it has been suggested that the solar dynamo stopped \cite{ref5,ref6}. The figure clearly shows the 11-years amplitude attenuated  during the Maunder, Dalton and Modern minima (negative phase of the periodicity of 120-years in Fig. 3c). The amplitude of the solar cycle never disappears completely, this is particularly so during the Maunder minimum \cite{ref38}.  After the year 2000 the peak amplitude of the cycle  tends to decrease, in fact during the 21st minimum ($2003$$-$$2063$) it is similar to the Dalton or Modern minima ($1883$$-$$1940$).

The phase also shows an  amplitude modulation, it presents an inversion  between $\sim$$1780$ and $1790$, as this in not a particularly maximum or minimum time, we suggest that it comes probably from a shortcoming of the TS-KRIVOVA reconstruction in this time-interval. 

From $2011$ to $2100$, the phase does not show any other inversion, probably indicating the good quality of our estimation. Fig. $3c$ shows the $120$-years cycle, we notice that it is the negative phase of this periodicity that coincides with the secular minima: Maunder  between the years $1637$$-$$1704$, Dalton between the years $1771$$-$$1829$ , Modern between  the years $1883$$-$$1940$ and the $21$st century minima between the years $2003$ and $2063$. 

\begin{figure}
\center
 \noindent\includegraphics[width=0.47\textwidth]{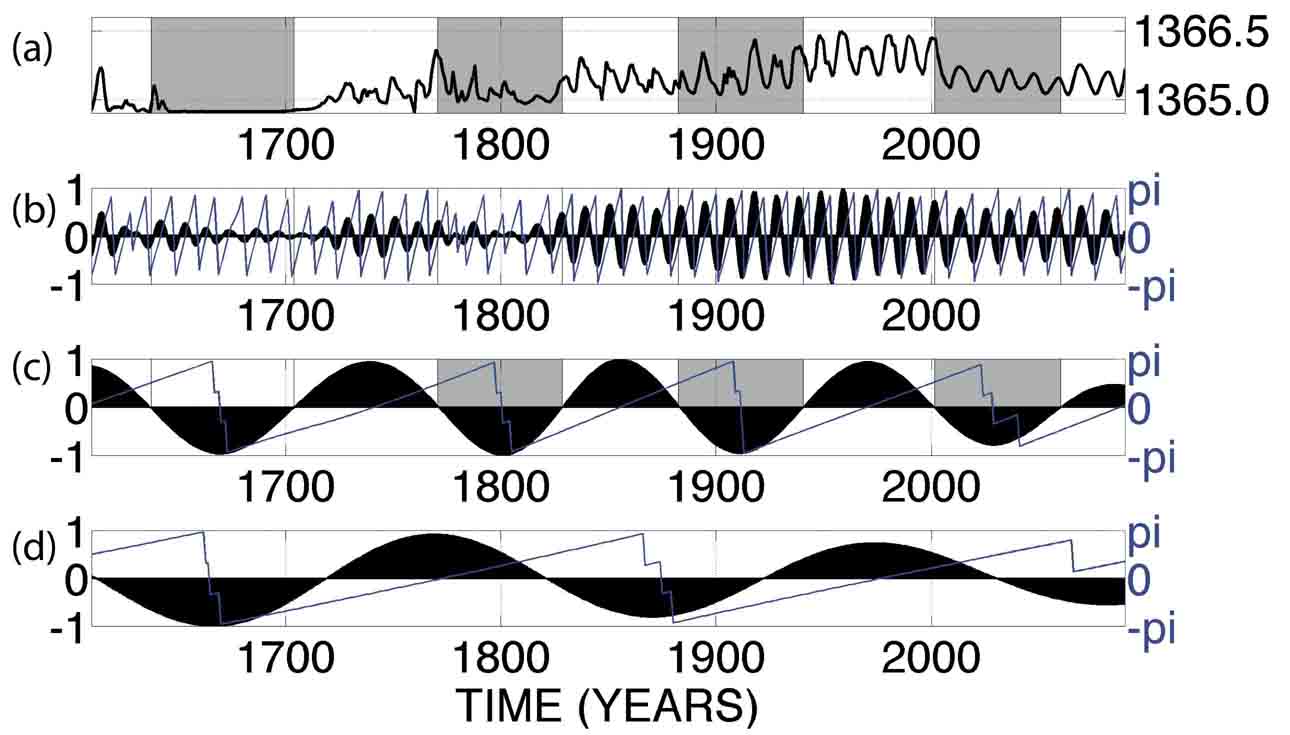}
 \caption{Phases (blue lines) and amplitudes (black areas) associated to the main periodicities of the  LS-SVM model. The light-gray shading indicates the time-span of the secular minima. (a) Modeled  time series . (b) The $11$-yrs cycle. (c) The $120$-yrs cycle. (d) The $240$-yrs cycle. }
 \end{figure}

Based on this periodicity we calculate the average TSI for the secular minima: Maunder $1364.8$ $Wm^{-2}$, Dalton $1365.2$ $Wm^{-2}$ Modern $1365.5$ $Wm^{-2}$ and the $21$ st century $1365.4$ $Wm^{-2}$ $\pm$ $1\sigma=0.4318$ $Wm^{-2}$, this TSI value is between the Dalton and Modern minima.  Again we observe that the phase does not change after $2010$.

The Modern maximum noticed between the years $1940$ and $2003$ has an average TSI of $1366$ $Wm^{-2}$,  and the $21$st century minimum has a TSI average of  $1365.4$ $Wm^{-2}$, this implies a negative radiative forcing of  $\sim$$-0.1$ $Wm^{-2}$. The radiative forcing for the Maunder minimum to the Modern maximum\cite{ref31} is $\sim$$0.21$ $Wm^{-2}$, then the radiative forcing  associated to the $21$st century minimum is almost half of that forcing.  

Finally, Fig. $3d$ shows the $240$-years cycle, it seem that according to the lag between the $120$-years and $240$-years cycles we have different amplitudes of the secular solar minima,  for instance, when its negative phase  coincides with the $120$-years cycle we have the deepest minima, like the Maunder minimum.  Again, the phase of this periodicity  holds for the next $100$ years.

To decide when the solar  activity is ``high" or ``low", we calculate the power of the TSI as a direct indicator of energy released by the solar dynamo and the level of activity for each solar cycle. We use the mean power value of the PMOD composite ($1976-2010$) to calculate  the anomalies for each cycle.  The power anomalies are normalized and appear as blue bars in  Fig. 4,  which also presents the TSI (solid line).  There are positive power anomalies during solar cycles $21$ and $22$, coinciding with the end of the Modern maximum. Between solar cycles $23$ and $30$ the power anomalies are negative, coinciding with negative phase of the $120$-years periodicity. This reinforces the result of Fig. 3c  suggesting  that this periodicity is closely  associated with secular minima. According to the power of the anomalies,  solar cycles $23$--$25$ and $30$ could present lower activity than cycles $26$ to $29$, regardless of the fact that the peaks of cycles 27 and 28 are the lowest. 
 
 \begin{figure}
\center
 \noindent\includegraphics[width=0.47\textwidth]{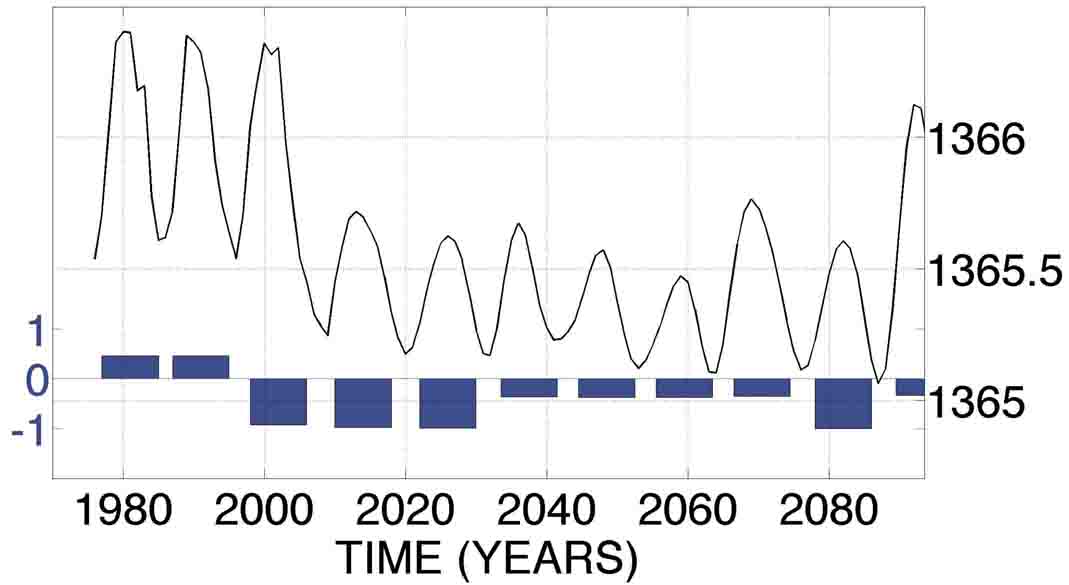}
 \caption{
  We use the mean power PMOD composite (1976-2010) to calculate  the power anomaly for each cycle. The power anomalies are normalized and appear as blue bars. The TSI is the solid line. Negative (or Positive) power anomalies is  `Low'' (or ``High'') solar   activity}
 \end{figure}
 
 The calculated power anomalies show that  low solar secular activity occurs when there are negative anomalies and  high solar secular activity  appears with positive anomalies. It is possible that the zero in the anomalies, represents the normal state of the dynamo. The ``Prolonged Sunspot Minimum''  discovered by Maunder, represents a phase of solar history and corresponds to a special state of the dynamo when it is working well below its average power.
\\\\
 This work was partially supported by DGAPA-UNAM, IN$103209-3$, IN$117009-3$, IN$105909$ and  IN$117009$ grants, IXTLI: IX$101010$ and IX$100810$ grants and CONACYT-F$282795$ grant. 

\nocite{*}
\bibliography{IGF}
\end{document}